\documentclass[prl,aps,twocolumn,showpacs,floatfix]{revtex4}
\usepackage{graphicx,amssymb, amsmath, natbib}
\usepackage{verbatim}
\usepackage{times}
\begin{document}

\def\MEMO#1 {}
\def\HIDE#1 {{}}
\def\OMIT#1 {{}}
\def\SAVE#1 {{}}
\newcommand{\prlsec}[1]{{\underline{\it #1} --- }}
\newcommand{\eqr}[1]{(\ref{#1})}
\newcommand{\vg} {{\vec{v}_g}}
\newcommand{\kk} {{\vec{k}}}
\newcommand{\pp} {{\vec{p}}}
\newcommand{\kkE} {{\kk_\epsilon}}
\newcommand{\kkR}{{\kk_\RR}}
\newcommand{\rr}{{\vec{r}}}
\newcommand{\RR}{{\vec{R}}}
\newcommand{\En}{{\omega}}
\newcommand{\imp}{{\rm imp}}
\newcommand{\Te}{{T_e}}               
\newcommand{\tmatrix}{{T}}            
\newcommand{\Gnonsing}{G_{\text{non-singular}}}
\newcommand{\ANO}{_{\text{ANO}}}
\newcommand{\ORD}{_{\text{ORD}}}

\title{Echolocation by Quasiparticles}
\author{Sumiran Pujari and C. L. Henley}

\affiliation{Department of Physics,
Cornell University, Ithaca, New York 14853-2501}

\begin{abstract}
\SAVE{With the advent of Scanning Tunneling Microscopy (STM), it is now possible to probe a 
material at a microscopic level that is unprecedented. This necessitates the important 
question: How much information can STM give us about the microscopic (spatially varying)
Hamiltonian of a material?} 
It is shown
that the local density of states (LDOS), measured in an Scanning Tunneling Microscopy (STM) experiment, at
a single tip position contains oscillations as a function of {\it energy}, due to
quasiparticle interference, which is related to the positions of
nearby scatterers. We propose a method of STM data analysis based on this idea, which 
can be used to \emph{locate} the scatterers.  In the case of a superconductor,
the method can potentially distinguish the nature of the scattering by a particular impurity.
\end{abstract}




\maketitle

Scanning Tunneling Microscopy (STM), which measures the ``local density of states'' (LDOS) 
as a function of position and energy set by the bias voltage, 
has opened the door to imaging the sub-nanoscale 
topography and electronic structure of materials,
including normal metals~\cite{Eigler} and especially cuprate superconductors
~\cite{Yazdani, Hudson, Pan, Cren, Howald, Lang, Hoffman, Renner}. 

The dispersion relations of (Landau or Bogoliubov) quasiparticles
may be extracted from STM data 
on normal metals~\cite{Sprunger-Petersen,wenderoth}
and superconductors~\cite{McElroy-Hanaguri},
via the inverse method called 
Fourier transform scanning tunneling spectroscopy 
(FT-STS)~\cite{Sprunger-Petersen,McElroy-Hanaguri},
or directly in real space~\cite{wenderoth}.
This technique is based on the fact that impurities produce 
spatial modulations of the LDOS in their vicinity -- standing
waves in the electronic structure that generalize the 
Friedel oscillations found in metals at the Fermi energy.
In the cuprates BSCCO and CaCuNaOCl~\cite{McElroy-Hanaguri},
experiments showed these quasiparticle oscillations 
were dominated by eight wavevectors that
connect the tips of ``banana'' shaped energy contours in reciprocal space,
the so-called Octet model as explained theoretically~\cite{WangLee-Ting}.
For optimally doped samples, the dispersion inferred from
these wavevectors agrees well with $d$-wave BCS theory indicating
the existence of well-defined BCS quasiparticles in this regime. 

\SAVE{
For under-doping and over-doping, additional nondispersing modulations are
seen in STM, having the symmetry of a charge order (but not necessarily
entailing a charge variation).
Their nature is still controversial and this paper does not 
discuss these features of STM phenomenology.}

The central observation of this paper is that the 
same Friedel-like oscillations of the LDOS,
analyzed in the space/momentum domain by FT-STS,
are also manifested in the energy/time domain.
Our analysis shows that the small impurity-dependent modulations 
of the LDOS have a period, in energy, inversely proportional 
to the time required by a quasiparticle wavepacket 
to travel to the nearby impurities and back -- hence  we call it
``quasiparticle echo''. From this, in principle,
one can determine the location and (in a superconductor)
the nature of the point scatterers in a particular sample.

\SAVE{We represent energy by $\omega$ not $E$ throughout.}

\begin{figure}
    \begin{tabular}{cc}
		a) \resizebox{40mm}{!}{\includegraphics{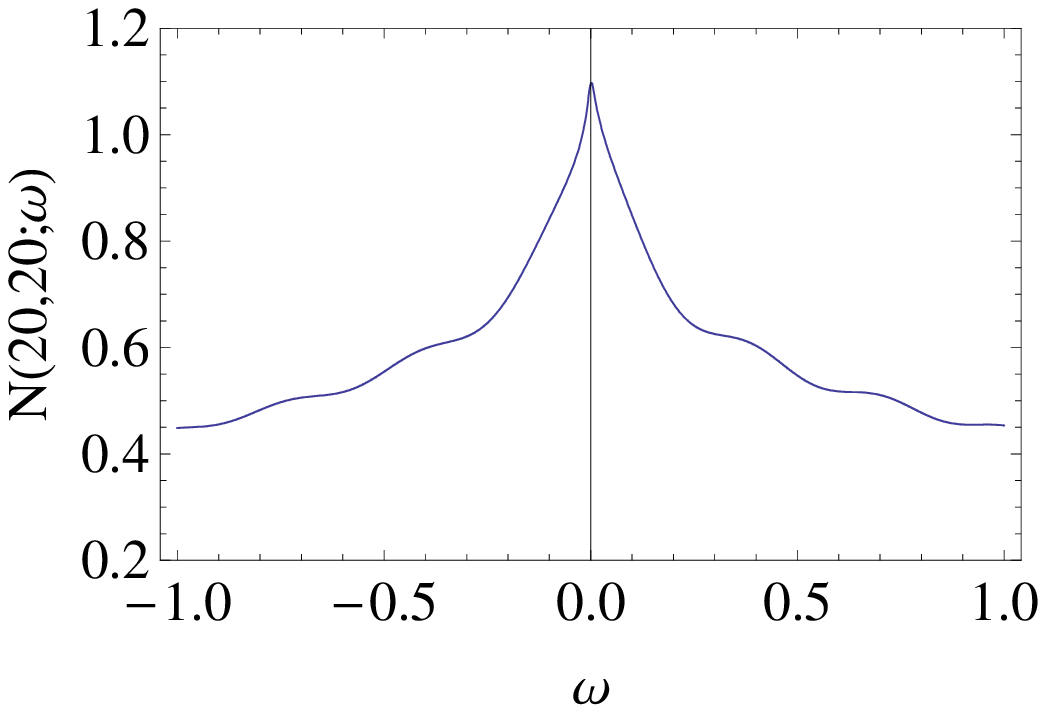}} & b) \resizebox{40mm}{!}{\includegraphics{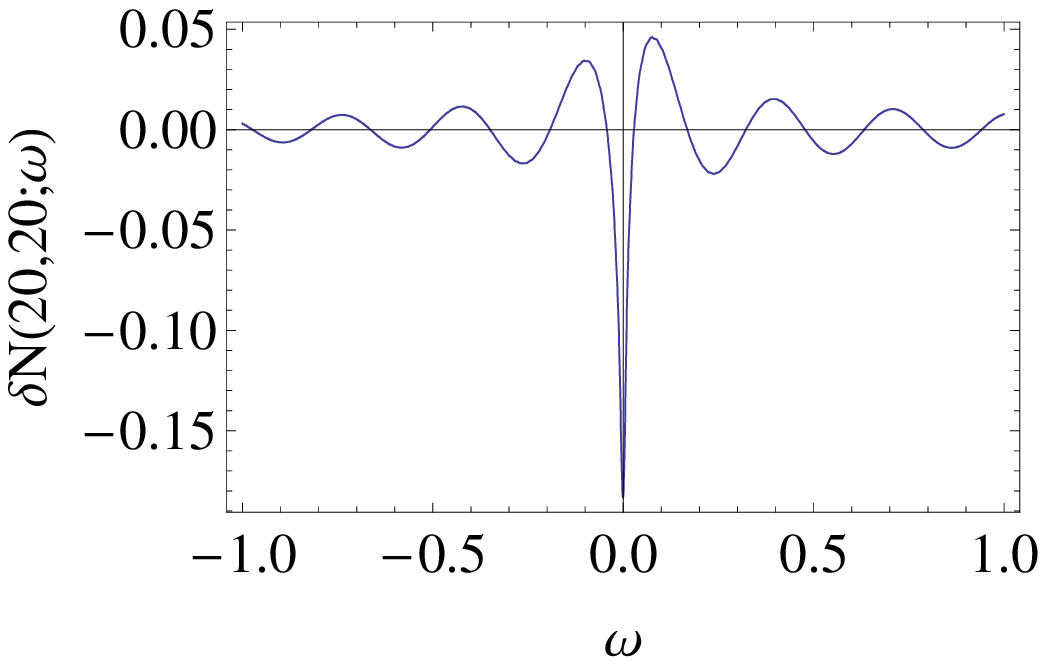}} \\
		c) \resizebox{40mm}{!}{\includegraphics{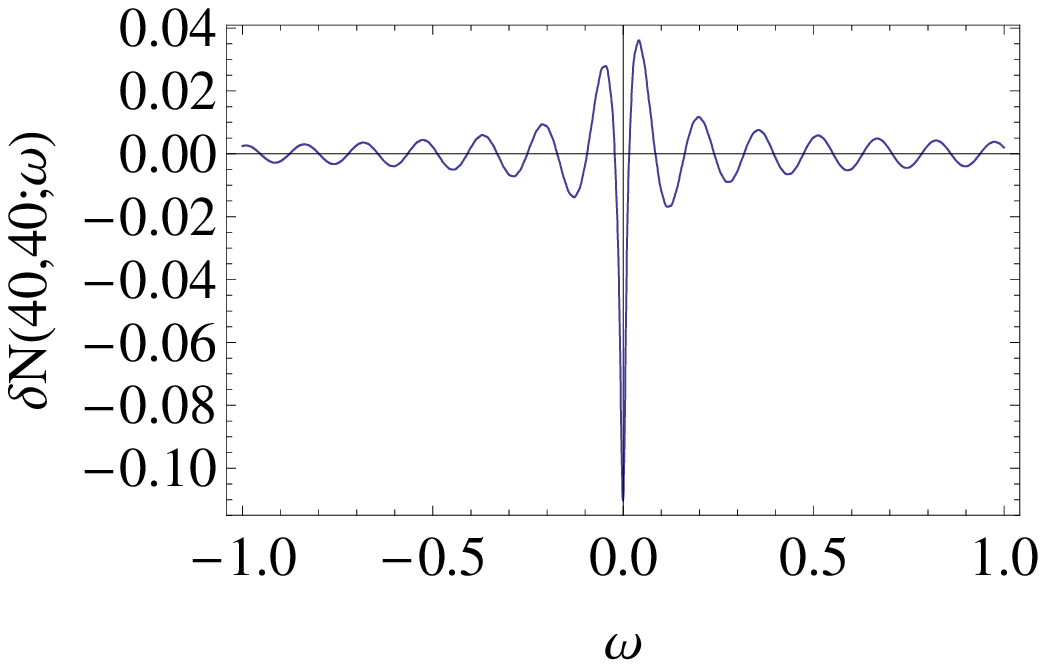}} & d) \resizebox{40mm}{!}{\includegraphics{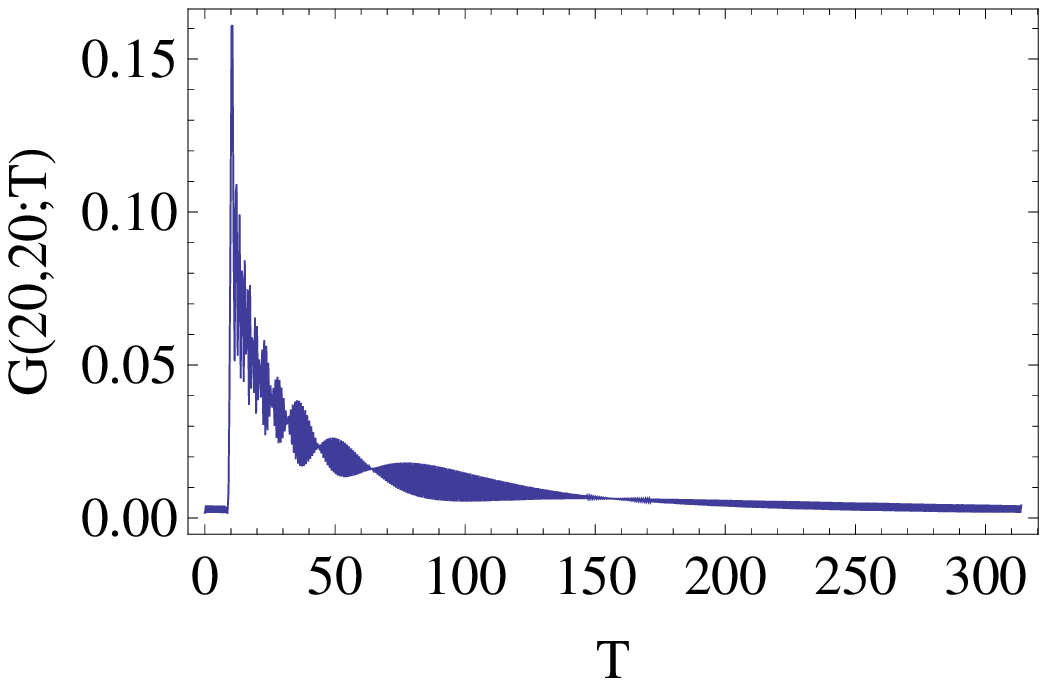}} \\
    \end{tabular}
    \caption{LDOS as a function of energy, showing oscillations due to quasiparticle echoes.  
(a) LDOS at a point $20\sqrt{2}$ away from a point impurity along the $[1,1]$ direction(lattice constant $ = 1$). 
(b) corresponding LDOS after subtracting the clean LDOS : $\delta N(20,20;\En)$, (c) $\delta N(40,40;\En)$, 
(d) Magnitude of Local Green's function as a function of time : $|G(20,20;T)|$. \MEMO{From CLH: is my rewrite true?}
The singularity appears at time $\Te/2$, where $\Te$ is given by    \eqr{eq:basic}.
\MEMO{to SAVE: There's no spectral weight below a time $\Te/2$.}
As we change the distance along this direction, the shortest echotime changes in proportion in accordance with our semiclassical
expectations.}
    \label{fig:norm_echoes}
\end{figure}

\prlsec{Quasiparticle echo}
The basic idea of the LDOS modulations may be understood semiclassically.
The LDOS $N(\rr;\En)$ is defined as $-(1/\pi) {\rm Im} G(\rr,\rr;\En)$,
the time Fourier transform of the local (retarded) Green's function $G(\rr,\rr;t)$.
Imagine a bare electron wavepacket (centered on energy $\En$)
is injected at time $t=0$ at point $\rr$ in a two-dimensional material:
the Green's function expresses its subsequent evolution.
Assuming there are well-defined quasiparticles at this energy with 
dispersion $E(\kk)$; then for every wavevector $\kk$ on the energy
contour $E(\kk)=\En$, the wavepacket has a component spreading
outwards at the group velocity $\vg(\kk)\equiv \nabla_\kk E(\kk)/\hbar$.
When this ring reaches an impurity at $\rr_\imp$, it  serves as a secondary 
source and the reflected wavepacket arrives at the ``echo time''
   \begin{equation}
     \Te\equiv 2 \frac{|\RR|}{|\vg(\kk)|}
   \label{eq:basic}
   \end{equation}
for the $\kk$ such that 
$\vg(\kk) \parallel \RR\equiv \rr_\imp-\rr$.  This creates a sharp peak 
at $t=\Te$ in $G(\rr,\rr;t)$ [see Fig.~\ref{fig:norm_echoes} (d)],
and hence modulations as a function of 
$\En$ in its Fourier transform $N(\rr;\En)$ with period 
$\Delta\En= 2\pi\hbar/\Te$~\cite{gutzwiller}.
Generically, for a particular impurity direction,
$|\vg|$ varies with energy,  so the the modulation in
$\delta N(\En)$ due to the impurity is ``chirped'' correspondingly.

We illustrate the quasiparticle echo first by a numerical calculation
for a normal metal, defined by the lattice Schrodinger equation
for the wavefunction $u_i$ on site $i$:
\begin{equation}
\sum_{j} (t_{ij} + \mu_i \delta_{ij})  u_i = E  u_i  .
\end{equation}
Here the $t_{ij}'s$ are intersite hoppings 
and the $\mu_{i}$'s are on-site potentials (including the chemical potential); 
in this paper, we assume they are
translationally invariant except at discrete (and dilute) impurity sites.
We take the specific case of nearest-neighbor hopping $t$ at half-filling, so the
the dispersion is $\epsilon(k_x,k_y)=-2t(\cos{k_x}+\cos{k_y})$, 
and we place one (repulsive site potential) impurity at the origin. 
To numerically calculate the LDOS, we used the Recursion method \cite{recursion},
which is well-suited for cases without translational symmetry. 

Fig. \ref{fig:norm_echoes}(a) shows the impurity case LDOS which has echo oscillations
on top of what otherwise would have been clean case LDOS, visible
along the sides of the peak.
\SAVE{Apart from the oscillations, the LDOS looks the same as the pure
(no impurity) case.}
\HIDE{[except the impurity makes the Van Hove singularity much less sharp,
producing a negative spike in the subtracted LDOS, Fig. \ref{fig:norm_echoes}(b,c)]}
Note that, for us to see more than one oscillation within the bandwidth, the impurity
must be at least several sites away; hence the oscillations always have small
amplitude and are best viewed by subtracting the clean LDOS. Throughout
the paper, energy is in units of $t$ and time in units of $t^{-1}$ with $t=1$
and $\hbar=1$.

For a given energy $\En$, we define $\Delta \En (\En)/2$
as the separation of the zeroes that bracket $\En$
in the (subtracted) $\delta N(\En)$ trace, and let $\Te(\En) \equiv 2 \pi\hbar/\Delta\En(\En)$.
We chose $E=0.7t$ and $\RR$ in the [1,1] direction, for which the group
velocity is $v_g= 2.785 t/\hbar$. Then, using $\delta N(20,20;\En)$,
$\delta N(30,30;\En)$, and $\delta N(40,40;\En)$ [the first and last trace of these
are shown in Fig.~\ref{fig:norm_echoes}(c,d)],
we read off $\Delta\En/2= 0.1545$, $0.103$, and $0.077$, from which $v_g T_e/2= 20.04\sqrt{2}$,
$30.05\sqrt{2}$, and $40.22\sqrt{2}$, respectively.
The proportionality between the oscillation rate and the actual distance
confirms the semiclassical explanation of these modulations.

\SAVE{$\vec{v}_{group}(k_x,k_y) = 2(\sin{k_x} \hat{x} + \sin{k_y} \hat{y})$. 
In general, 
$\hbar|\vec{v}_{group}(1,1)|=2\sqrt{\sin^2{k_x}+\sin^2{k_y}}$.
Along $(1,1)$, $\cos{k_x}=\cos{k_y}$ and $\sin{k_x} = \sin{k_y}$,
so $\hbar|\vec{v}_{group}(1,1)|= 2\sqrt{2(1-\cos^2{k_x})} 
= 2\sqrt{2(1-(-E/4)^2)} = 2.785$.}

{\prlsec{Echolocation}}
Using these quasiparticle echoes, we can locate the position of impurities by measuring 
the LDOS wiggles at a few points in the vicinity.
At each point, we extract the wiggle period $\Delta \En$ 
and hence the echo time $\Te\equiv 2\pi/\Delta\En$.
Then \eqr{eq:basic}
defines a locus of possible impurity locations, 
$\{\vec{v}_{group}(\kk) \Te/2: \epsilon(\kk)=\En$.
The intersection of the loci from STM spectra taken at
multiple points $\rr$ will locate  $\rr_\imp$ uniquely.
Furthermore, via a more exact derivation of the LDOS
modulations (see below), the amplitude of the LDOS modulations
tells the scattering strength of the impurities
(in Born approximation they are proportional to each other). 
Once an impurity has been pin-pointed, 
the higher-energy STM spectrum at that point
may independently identify the chemical nature
of the impurity, e.g. in cuprates~\cite{McElroy-chem}
and thus may reveal which  kinds of impurities 
are important for the scattering of quasiparticles.

\begin{figure}
    \begin{tabular}{cc}
      a) \resizebox{40mm}{!}{\includegraphics{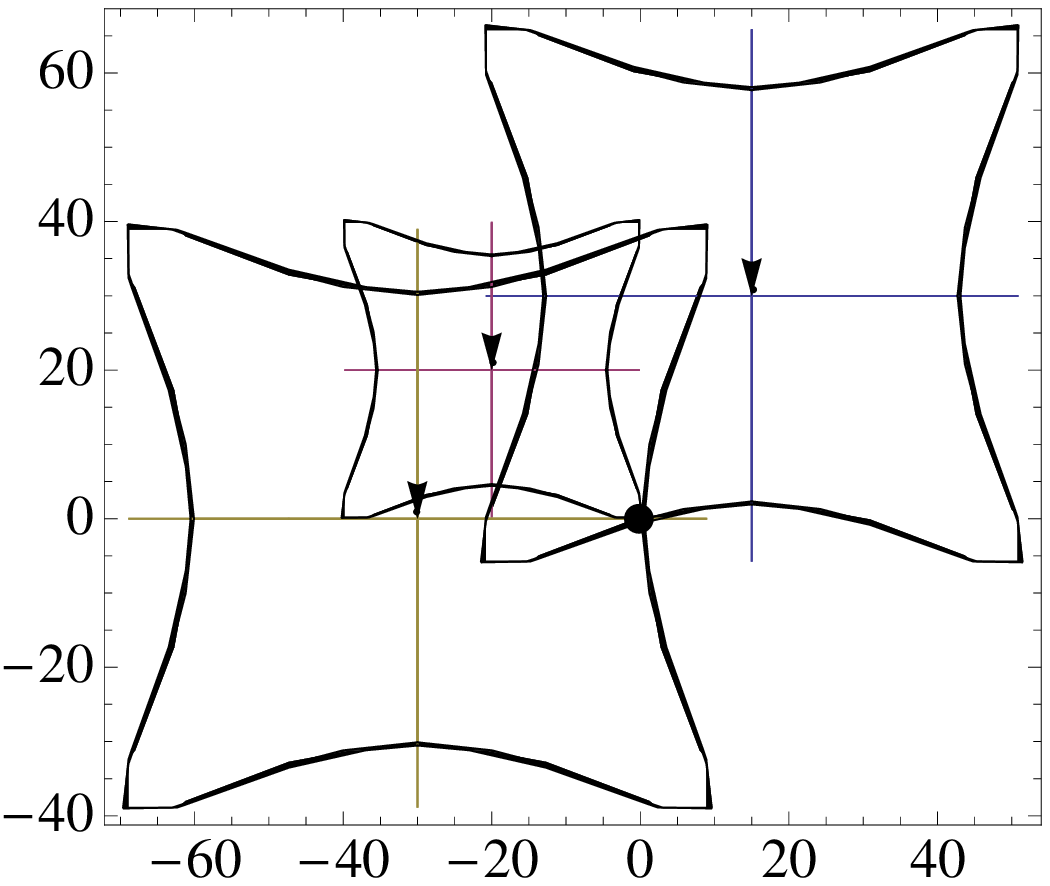}} & b) \resizebox{40mm}{!}{\includegraphics{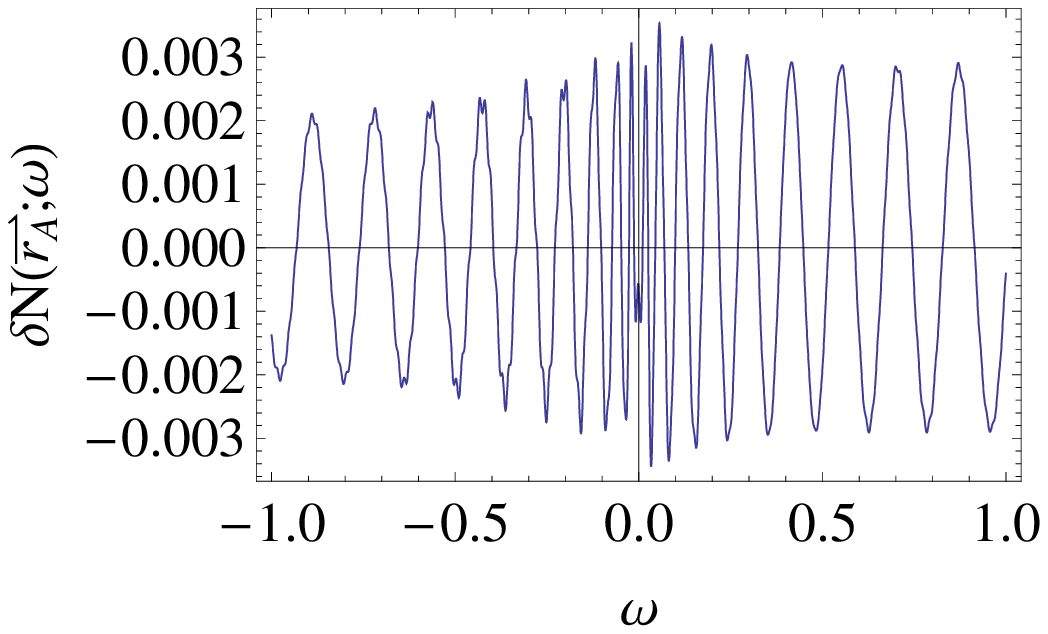}}  \\
      c) \resizebox{40mm}{!}{\includegraphics{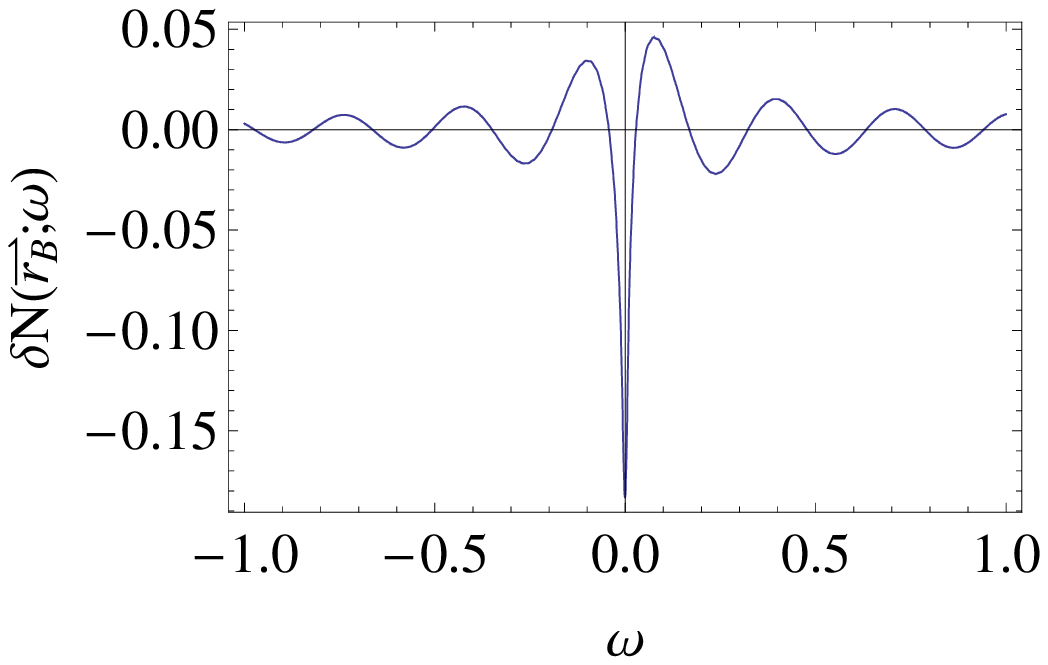}} & d) \resizebox{40mm}{!}{\includegraphics{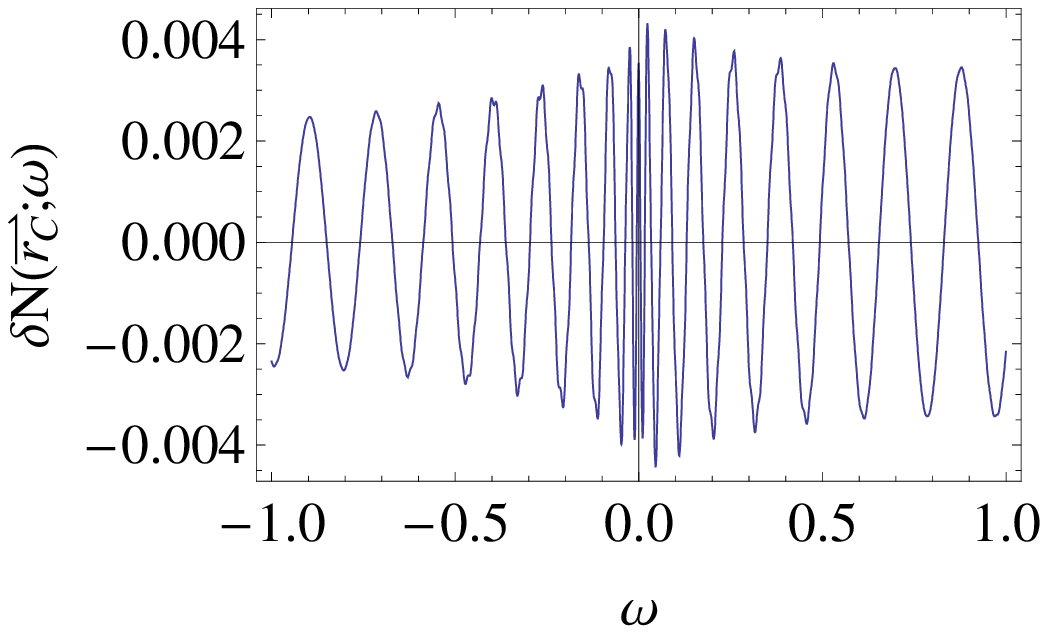}} \\
    \end{tabular}
    \caption{(a) Schematic of few measurements around an impurity. The arrowheads represent the STM Tip postions.
 After measurement, we get (b) $\delta N(\rr_A;\En)$,
(c) $\delta N(\rr_B;\En)$ and (d) $\delta N(\rr_C;\En)$. Extracting the echotimes for each measurement at $\En = 0.7t$, we locate
the impurity, shown as a black dot, in the first panel.
Note, that the locus of impurity locations changes with $\En$, and is of the shape shown only at $\En = 0.7t$.}
    \label{fig:echodata}
\end{figure}

As a test, we evaluated the subtracted LDOS at three points
$\rr_A=(-30,0)$, $\rr_B=(-20,20)$, and $\rr_C=(15,30)$,
with the impurity at $\rr=0$.
From the half-periods of the wiggles at energy $=0.7 t$, (extracted as before)
we found the respective echo times
$T_A=39.9$, $T_B=20.4$ and $T_C=36.7$.
\MEMO{Should these numbers be moved to figure caption? Sumi: OK with either.}
The three scaled loci(scaled by half the respective echotimes), shown in Fig. \ref{fig:echodata} e),
intersect at (0,0) as can be seen graphically, thereby demonstrating the idea of echolocation.
A more careful numerical analysis can be done
to extract errors in echolocation as well.
\MEMO{To do: actually FIND IT NUMERICALLY (so as to have an error.)}

\MEMO{CLH: Implicitly, talking of a unique $\Delta E$ assumes
a constant $\vg$... if $|\vg|$ varies that is just a 
complication but in principle, it is actually more
information... Say this in a footnote?
Sumi: YES and we refer to it
the first time we do the energy selective analysis.
CLH (NEW): Just where is that, can't find it.}

\prlsec{Analytic derivation}
Adopting the T-matrix formalism,
we can obtain an analytic form for the LDOS modulations.
Formally, the difference in dirty LDOS and clean LDOS for a single point impurity is given by 
   \begin{equation}
   \delta N(\rr;\En) = -\frac{1}{\pi} {\rm Im}
   \Big[G_0(\rr-\rr_\imp;\En) \tmatrix(\En) G_0(\rr_\imp-\rr;\En) \Big]
    \label{eq:deltaNG0}
\end{equation}
where $G_0(\rr,\rr_\imp;\En) \equiv G_0(\rr-\rr_\imp;\En) \equiv G_0(\RR;\En)$
is the free propagator;
LDOS modulations are  due to interference between the two $G_0$ factors.
  \begin{equation}
  G_0(\RR;\En) 
  = \lim_{\delta \rightarrow 0^+} \int_{B.Z.} \frac{d k_x d k_y} {(2\pi)^2} 
          \frac{e^{i \kk. \RR}}{\En+i\delta-\epsilon(\kk)}
  \label{eq:G0BZ}
  \end{equation}
\MEMO{Sumi: would move the algebra explanation (next para) a long
footnote,  save space w/smaller fontsiz?  CLH: No, this seems pretty
central to the derivation of the formulas displayed.}
The integrand is singular all along the energy contour
$\epsilon(\kk)=\En$, which we also parametrize as $\kkE(s)$, 
where $s$ is the arc-length in reciprocal space.
\SAVE{($\kkE$ implicitly depends on $\En$ too.)}
By the change of variables $z\equiv e^{ik_y}$
we convert the inner ($k_y$) integral to a complex contour integral
in the $z$ plane (rewriting $\epsilon(k_x,k_y)$ 
as an analytic function of $z$); 
for $k_x$ values found on the energy contour, the $z$ path encounters
two poles, one inside and one outside, depending on the sign of $\delta$.
Extracting the residue and absorbing factors, we get
\begin{equation}
  G_0(\RR;\En) 
 = \frac{1}{2\pi i}  \oint
\eta(s) ds 
     \frac{e^{i \kkE(s)\cdot\RR}}{\hbar v_g(\kkE(s))} +\Gnonsing
\end{equation}
where $\eta(s) =1$ on the half of the energy contour where
${\mathrm{sgn}(\delta)=\mathrm{sgn}(|\vg (\En, s)|)}$ and zero on the other half.
\OMIT{also, $v_g (\hat{R}, \En) = |\nabla \epsilon(\kkR(\RR,\En)|$.}
The non-singular term $\Gnonsing$ comes from the integrals over $k_y$ which 
do {\it not} cross the energy contour.

At large $\RR$, 
the two-dimensional BZ integration will be dominated by those $\kk$
\cite{FN-contour-sing}
on the energy contour where the phase in the numerator is stationary,
i.e. $\vg(\kk) \parallel \RR$: let us call such a point $\kkR$
(so it is a function of the direction $\hat{R}$ and of $\En$). 
Using standard formulas of the stationary phase approximation~\cite{asymptotic}
we get asymptotically
   \begin{equation}
   \label{eq:statphase}
G_0(\RR;\En) = \frac{-i e^{i \pi/4}}{v_g}
                \sqrt{\frac{1}{2 \pi \kappa |\RR|}} e^{i\kkR(\RR,\En)\cdot\RR}.
   \end{equation}
Here $\kappa^{-1}$ is the curvature $d^2\kkE/ds^2$ of the energy contour at $\kkR$.

\SAVE{Note thet the magnitude of quasiparticle interferences 
is determined by an interplay of 
the magnitude of group velocity and that of the curvature of the contour. 
Generally, portions of high curvature in a contour correspond 
to low group velocity and vice versa; 
hence there is a tendency for these factors to cancel.}

Using \eqr{eq:basic} and \eqr{eq:deltaNG0}, we finally get
  \begin{equation}
   \delta N(\En) = \frac{\tmatrix} {2 \pi^2 v_g^2 \kappa R} 
    \cos\Big(2\kkR(\RR,\En ) \cdot \RR\Big).
   \label{eq:dNresult}
   \end{equation}
valid  in the limit of a distant impurity.
(All factors are actually functions  of $\RR$ and $\En$: 
these arguments are shown only in the rapidly varying factors.)
As we change $\En$ to $\En + \delta \En$ keeping $\RR$ fixed, 
the chain rule gives
$\kkR(\En + \delta \En) -\kkR(\RR,\En ) = v_g^{-1} \delta \En \hat{R}$
\SAVE {The intermediate step is $\delta \En d \kkR(\RR,\En )/d \En.$}
so, with $\phi = \kkR \cdot \RR$, we get 
   \begin{equation}
    \cos \Big(2\kkR(\RR,\En +\delta \En ) \cdot \RR\Big) \to \cos 
     (\phi + \Te \delta \En).
   \label{eq:dNresult-energy}
   \end{equation}
This confirms the simple semiclassical prediction $\Delta\En=2\pi/\Te$ 
(see Eq.~\eqr{eq:basic}) for the modulation period due to echoes.
The same quasiparticle interference is responsible for the
spatial oscillations evident in 
\eqr{eq:dNresult} and the  energy oscillations in 
\eqr{eq:dNresult-energy}.


\SAVE{In the above, we assumed
that the prefactor $L(\RR; \En)$ does not vary much with $\En$.}

\SAVE{The salient features of this demonstration 
of echoes are : 1) we see that the below a certain time,
there are no Fourier components implying $\delta N$ contains contribution above that time, 2) This time
is the shortest echotime and it, therefore, corresponds to the fastest group velocity, 3) This corroborates
with the fact that the weight for shortest echotime/fastest group velocity is highest, since along (1,1) direction
the fastest group velocity comes from the middle of the band($\En = 0$) which is where the VanHove singularity sits. 
This is also seen in the spikes sitting at the same energy in Fig. \ref{fig:norm_echoes}.}

\begin{figure}[h]
    \begin{tabular}{ll}
		a) \resizebox{40mm}{!}{\includegraphics{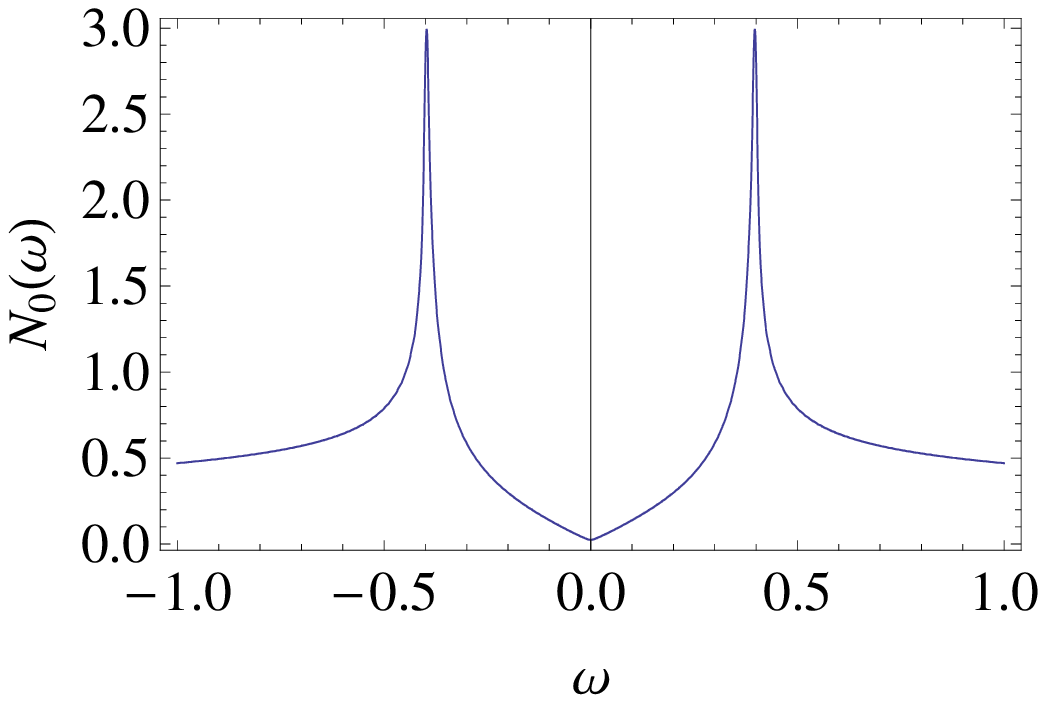}} & b) \resizebox{35mm}{!}{\includegraphics{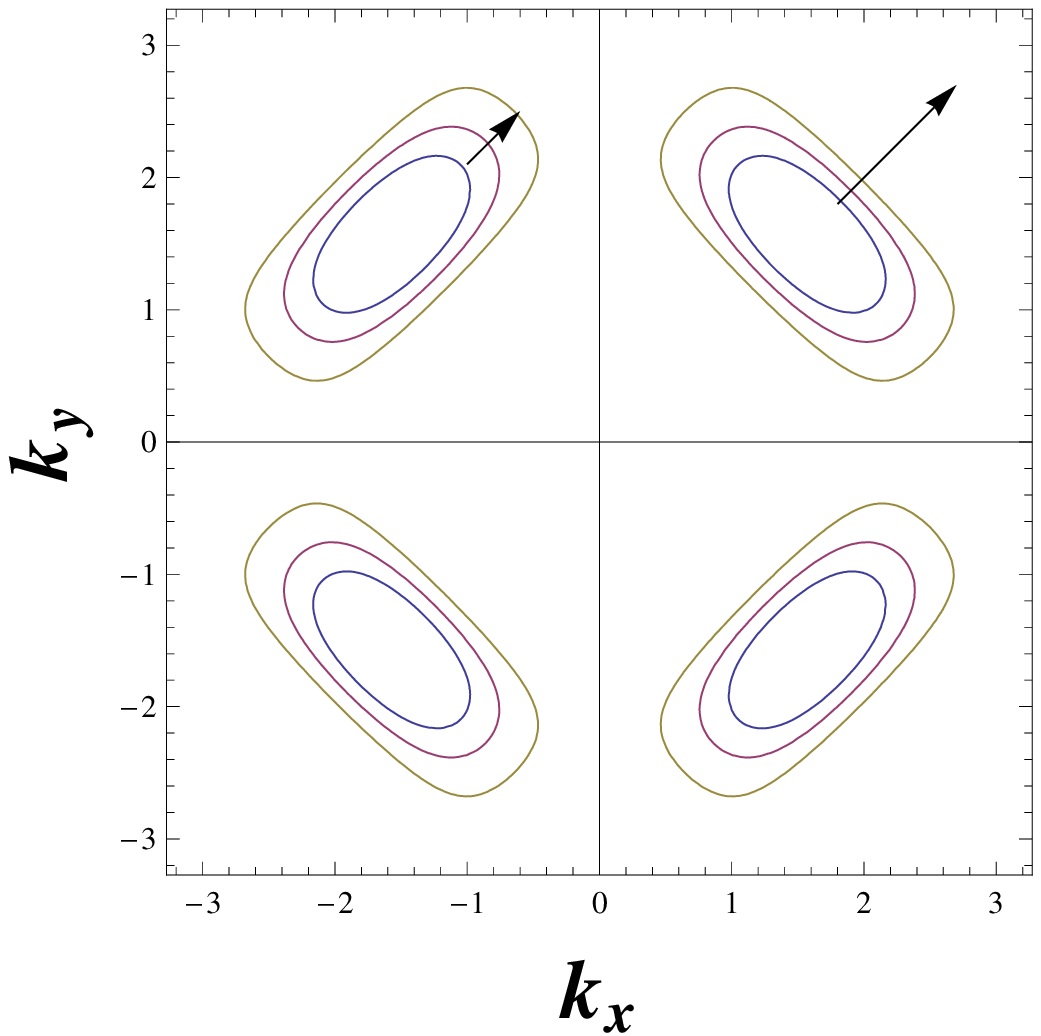}} \\
		c) \resizebox{40mm}{!}{\includegraphics{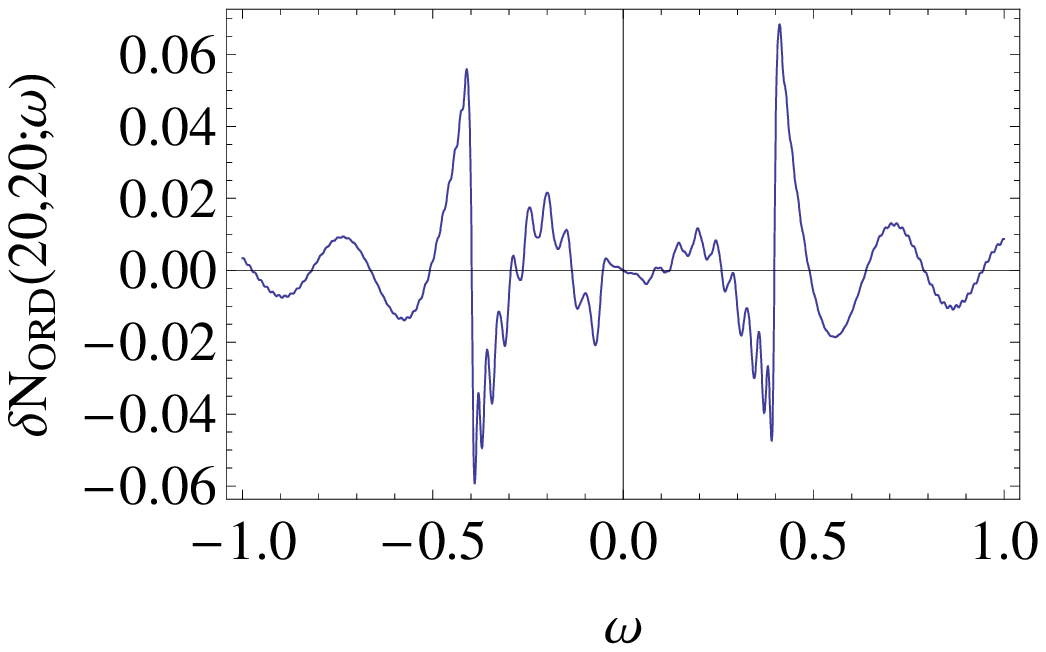}} & d) \resizebox{40mm}{!}{\includegraphics{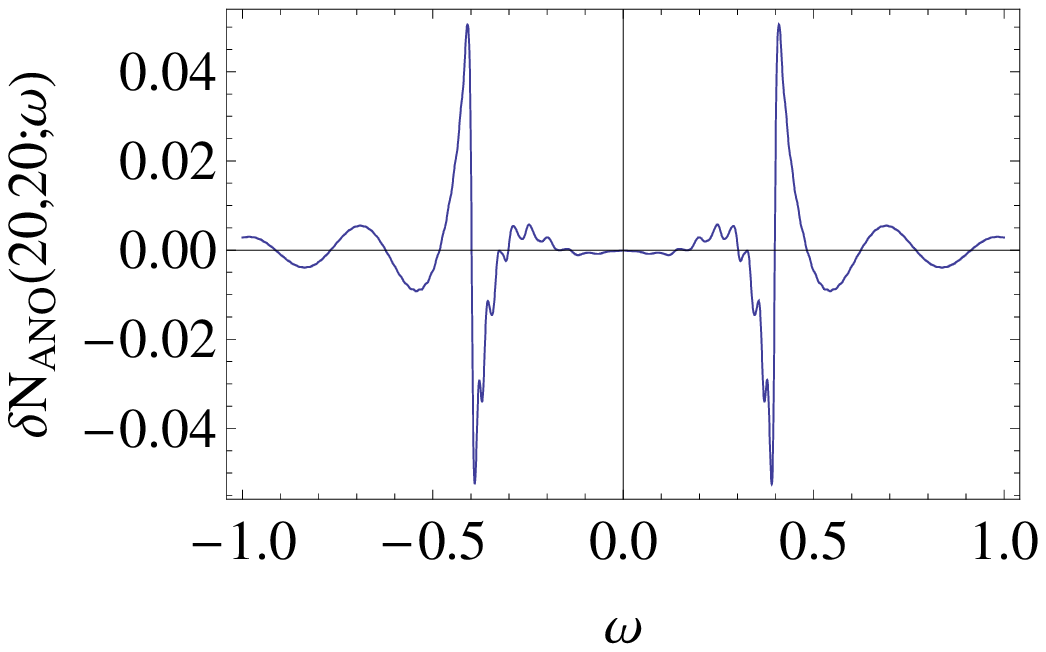}} \\
    \end{tabular}
    \caption{Quasiparticle echoes in a $d$-wave SC. (a) no impurity $N_0(\En)$ showing the d-wave gap,
(b) A caricature of the two different group velocities along (1,1) direction for d-wave Bogoliubov dispersion, 
(c) $\delta N_{\text{ORD}}(20,20;\En)$ for an ordinary impurity and (d) $\delta N_{\text{ANO}}(20,20;\En)$ for an
anomalous impurity.}
    \label{fig:SC_echoes}
\end{figure}

\prlsec{Echoes in cuprate superconductors}
Additional 
relevant issues arise in case of 
superconductors. To discuss these, we use a mean-field 
Bogoliubov-DeGennes(BDG) Hamiltonian with/without a single point impurity as shown
below.
\begin{equation}
\sum_{j} 
\left[ \begin{array} {lr}
t_{ij} + \mu_i \delta_{ij} & \Delta_{ij}\\
\Delta_{ij}^* & - t_{ij} - \mu_i \delta_{ij}
\end{array} \right] \left[ \begin{array}{c} u_i \\ v_i \end{array} \right]
= E \left[ \begin{array}{c} u_i \\ v_i \end{array} \right]
\label{BDG}
\end{equation}
where we are using a lattice formulation of BDG equations. The $u_i$s and $v_i$s represent particle and hole 
amplitudes on site $i$, $t_{ij}$s and $\mu_{i}$s represent the intersite hoppings and site chemical potentials
respectively, and $\Delta_{ij}$ represent the off-diagonal order parameter amplitude. We 
discuss $d$-wave  superconductors (dSC's)
to highlight this method's application to cuprates. 
For  dSCs, $\Delta_{ij}$ is nonzero only on nearest-neighbor bonds and 
$\Delta_{\hat{i},\hat{i}\pm\hat{x}}= -\Delta_{\hat{i},\hat{i}\pm\hat{y}}$ 
because of the $d$-wave nature. Our normal state 
is the same nearest neighbor tight binding model on the square lattice with $t = 1$ and 
off-diagonal hopping amplitudes set to $|\Delta|=0.1$. The Recursion method was
extended to superconductors in \cite{Litak} and is used for our numerics. 
\SAVE{In Fig. \ref{fig:SC_echoes} a), we 
show the LDOS for the clean case. One can see the d-wave gap clearly.} 
In Fig. \ref{fig:SC_echoes} c) and d), 
we show the LDOS(after subtracting the clean LDOS shown in Fig. \ref{fig:SC_echoes} a)) at $20\sqrt{2}$  
distance from an impurity along the $(1,1)$ direction  for the case of a potential scatterer 
and an anomalous pair potential scatterer (which scatters an electron into
a hole and vice versa) respectively.

In contrast to the normal case, there are two different 
wiggles : a fast one and a slow one. The reason for this is that the dSC
quasiparticle dispersion gives rise to two different group velocities in the $(1,1)$ direction \cite{11}. 
We also note that the fast wiggles exist only within the gap while the slow wiggles are 
both inside and outside the gap. \SAVE{Inside the gap, the fast and slow wiggles superimpose 
to give the resultant data.} In Fig. \ref{fig:SC_echoes} b),
we show the constant energy contours for the quasiparticle dispersion given 
by $E(\vec{k})= \sqrt{\epsilon(\vec{k})^2 + \Delta(\vec{k})^2}$, the gradient of
which is the quasiparticle group velocity. From 
Fig. \ref{fig:SC_echoes} b), we see that along $(1,1)$, the banana-shaped
energy  contours in the first and  third
quadrants give one velocity (which corresponds to the slow wiggles), 
while the contours in the second  and fourth quadrants give a slower 
velocity (which 
corresponds to the fast wiggles). 
For $E> |\Delta|$, there are no longer ``banana'' contours, 
so we get only one group velocity (similar to the normal case) and hence only 
one kind of wiggle is seen in Fig.~\ref{fig:SC_echoes}(c,d) outside the cusps.

Once the impurity is \emph{located} using the loci intersection method desribed before, 
one can study the LDOS data around the impurity to infer the
impurity's strength and  
whether it is ordinary (magnetic/nonmagnetic) 
(cf. Ref.~\onlinecite{Balatsky_review} and references therein)
or anomalous~\cite{Nunner}. 
This distinction is already visible in individual spectra:
provided the normal state is particle-hole symmetric, one gets
particle-hole {\it symmetric} echo oscillations 
$\delta N_{\text{ANO}}$ from an anomalous impurity, since it
scatters electrons into holes and vice versa [Fig.~\ref{fig:SC_echoes}(d)];
this is {\it not} the case for $\delta N_{\text{ORD}}$  
from an ordinary impurity
[Fig.~\ref{fig:SC_echoes}(c)].
\OMIT{This difference can act as a diagnostic for characterizing
impurities through their $\delta N$ signatures.}

A second diagnostic distingushing (nonmagnetic) ordinary scatterers from anomalous ones
is the real-space pattern of the surrounding standing waves in the LDOS,
which is best seen in Born Approximation. In this limit, the impurity T-matrix
is of the form (in the $2\times 2$ Nambu notation) 
$U_\imp\tau_3$ or $\Delta_\imp \tau_1$ for the ordinary or anomalous cases, respectively.
Then the echo oscillations take the respective forms
  \begin{equation}
    \delta N\ORD \propto U_\imp (G_{11}^2-G_{12}^2); \qquad  
    \delta N\ANO \propto \Delta_\imp (2 G_{11} G_{12}). 
  \end{equation} 
Here, the $G_{ij}$s are the matrix elements of the usual free propagator 
\OMIT{in Nambu notation}, 
$G_0(\kk; \En) = (\En^2-E(\kk)^2)^{-1} 
\big[\En  + \epsilon(\kk) \tau_3 + \Delta(\kk) \tau_1\big]$
thus in real space
\begin{equation}
   \label{eq:G_0SC}
    G_0(\RR; \En) = \frac{\pi i}{(2\pi)^2} \oint 
            \eta(s) \frac{ds}{2} g(\kk(s,\En)) + \Gnonsing
\end{equation}
where 
   $g(\pp;\RR,\En)\equiv 1+ 
   \frac{1}{\En} (\epsilon(\pp) \tau_3 + \Delta(\pp) \tau_1 )$.
\SAVE{Upon carrying out the stationary phase approximation for large $\RR$, we get
\begin{multline}
G_0(\RR; \En) = \frac{i e^{i \pi/4}}{2(2\pi)^2} \sqrt{\frac{2\pi}{\RR}} 
       \sum_{i} L_i g(\kk_i; \RR,\En) \cos(\kk_i\cdot\RR) 
\label{eq:G-nambu-final}
\end{multline}
where $L_i \equiv L_i(\RR;\En) \equiv  
|\nabla \epsilon(\kk)|^{-1} \;
|d^2 \kk(s,\En)/ds^2|_{\kk_i(\RR,\En)}^{-1/2}$;
the sum in \eqr{eq:G-nambu-final} is over all points $\kk_i$ on the energy contour
satisfying $\vg(\kk_i)\parallel \RR$. ($\kk_i$ implicitly depends on $\RR$ and $\En$.)
\MEMO{NEW to Sumi: Previous draft had inverse of an inverse, I assume this is right.
Also, I have taken a square root in $L_i$ def'n, need to check if that was correct.}
}

\begin{figure}[h]
    \begin{tabular}{cc}
     a) \resizebox{40mm}{!}{\includegraphics{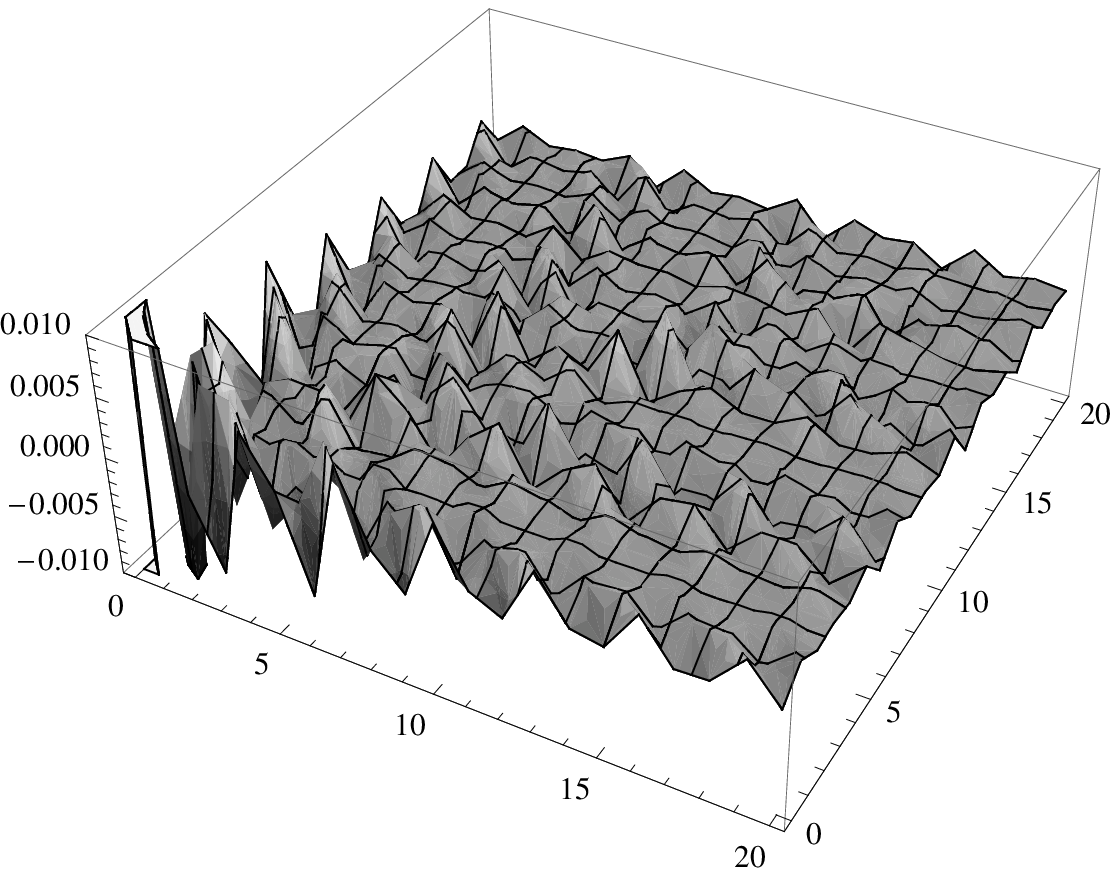}} & b) \resizebox{40mm}{!}{\includegraphics{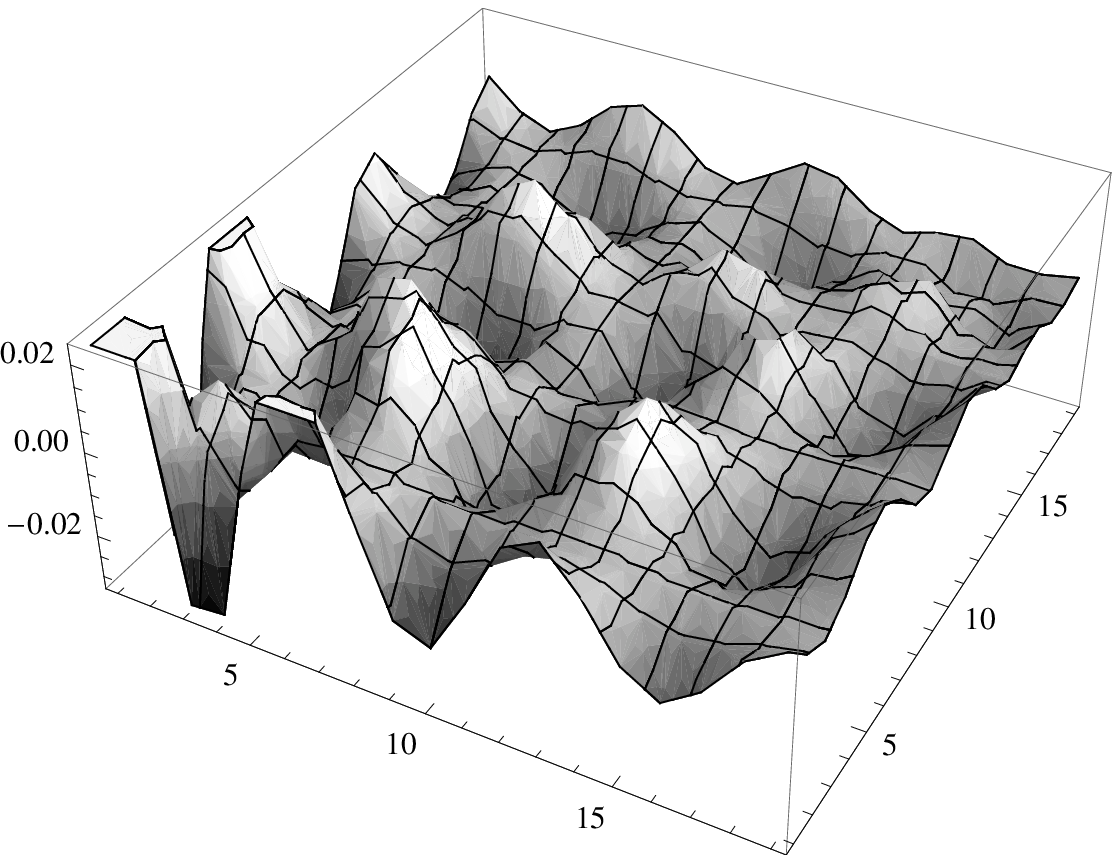}} \\
    \end{tabular}
    \caption{Shown are the $\delta N(\RR;\En=0.35t)$ around an impurity over a grid of (0,20)x(0,20) with other
quadrants related by symmetry. (a) $\delta N_{\text{ORD}}$, (b) $\delta N_{\text{ANO}}$. A subgap value of $\En=0.35t$ was chosen
arbitrarily.}
    \label{fig:SC_arndimp}
\end{figure}

We can carry out the stationary phase approximation as before, but instead we numerically calculated the propagator using 
Eq.~\eqr{eq:G_0SC}, since we are interested in LDOS information around(close) to the impurity. In  Fig. \ref{fig:SC_arndimp}, we show 
$\delta N$ around an impurity over a grid of 20x20 lattice points(shown one quadrant with
others related by symmetry).

We see that certain of the real-space oscillations, 
present in case of the ordinary impurity, are suppressed
in the case of a d-wave anomalous impurity.
This is the same effect as the suppression of certain ``octet" vectors
\cite{McElroy-Hanaguri,WangLee-Ting} for the case of d-wave anomalous impurity as argued in \cite{Nunner}'s Eq. 10 and the following paragraph.
\HIDE{
Note also in Fig. \ref{fig:SC_arndimp} that the magnitude/scale of oscillations in the d-wave anomalous impurity is larger. This is, presumably 
\MEMO{This claim needs to be numerically verified?}, the same as redistribution of weights
from "octet" peaks corresponding to suppressed oscillations to the unsuppressed ones as 
mentioned in the last line of the \cite{Nunner}'s above-quoted paragraph. 
\MEMO{Cite more exactly}
}
Our real-space analysis qualitatively duplicates that of Ref.\onlinecite{Nunner}
\MEMO{was theirs a real-space analysis?}, illustrating how
the real-space QPI  and our energy-domain echoes are complementary 
manifestations of the same phenomenon.

\HIDE{\MEMO{Keep this in the manuscript? trim it for now}
As we move to
higher impurity strengths, we see that the effects due to the two impurities are quite similar(figures not shown). 
A heuristic reason for this decrease
in the difference is as follows : for a point impurity, the T-matrix is $T(\En)= [1 - U.G(\RR=0;\En)]^{-1} . U$. In the Born Limit,
i.e., $U \rightarrow 0$, we can forget the $ U.G(\RR=0;\En)$ term in the bracket. Thus, $T \approx U$. While as we keep increasing the
impurity strength, gradually the $ U.G(\RR=0;\En)$ term will dominate over $1$. Thus, $T \approx G(\RR=0;\En)^{-1}$ which
is independent of what kind of impurity we have. Another point of note is that in the case of the s-wave anomalous
impurity, $\delta N$ is not symmetric about the (1,1) direction unlike for the ordinary case. The reason is a s-wave
scatterer in a d-wave background will distinguish between directions related by mirror symmetry along (1,1), while an 
ordinary or d-wave anomalous impurity does not.
}


\prlsec{Conclusion and Discussion}
In conclusion, we have introduced a method of STM data analysis
in the energy domain as
a phenomenological tool for the study  of real materials, 
complementary to FT-STS.
\SAVE{(in focusing on the energy/time dependence in
real space, rather than spatial Fourier transforms)}.  
Since it is based on the {\it same} quasiparticle interference effects 
already used successfully in FT-STS, we have confidence that the signals
will be observable.  They should be   
particularly strong in materials with an energy-dependent group velocity
in some range of energies, such as d-wave superconductors and also
graphene~\cite{graphene-STM}. 

\SAVE{In this paper, we illustrated echolocation only for the case of {\it isolated}
point-like impurities. Real STM data will typically contain a superposition
of LDOS wiggles from several nearby impurities; it is necessary (and not too
difficult) for the analysis method to separate these contributions.}

Since the echo  
analysis can be done in local patches of the sample (unlike FT-STS which
fourier transforms over a larger region), we can {\it locally}
verify the existence of quasiparticles at various energies through QPI.
In particular, in cuprates, echoes might be used to check the hypothesis
of quasiparticle extinction \cite{Kohsaka} above a certain energy.
Furthermore, we have argued that echo analysis might reveal the nature of
specific impurities \cite{pointimp} in a sample, information which hitherto was (at best)
known statistically.

\SAVE{In real materials, one can have
extended scatterers or chains of point impurities and these will have their own signatures
in an echo analysis. Analysing these will be a numerically-harder task and this will
form part of future work.}
\SAVE{Echoes rely on existence of well-defined Landau/BCS quasiparticles and thus mismatch of
this kind of analysis with experiment may point to features which are outside
a quasiparticle description and need different physics to be described. }

\OMIT{
(Alternately said)
It gives us confidence as to experimentally seeing echoes since, like two faces of the same coin, 
that real-space QPI  and 
energy-domain echoes are complementary manifestation of the same phenomenon.}

\HIDE{
We found that different kinds of
impurities have different experimentally measureable $\delta N$ 
signatures in LDOS.
\MEMO{``For echoes, it will mean no 
wiggles after a certain energy.''
CLH: I don't understand what that was about. 
WHAT means no echoes after a certain energy? Why?}}

\SAVE{Since the echo  analysis can be done in local patches of the sample (unlike FT-STS which 
fourier transforms over a larger region), we can {\it locally} 
verify the existence of quasiparticles at various energies through QPI. 
In particular, in cuprates, echoes might be used to check the hypothesis 
of quasiparticle extinction \cite{Kohsaka} above a certain energy. 
Furthermore, we have argued that echo analysis might reveal the nature of 
specific impurities in a sample, information which hitherto was (at best)
known statistically.}

\SAVE{In this paper, we illustrated echolocation only for the case of {\it isolated}
point-like impurities. Real STM data will typically contain a superposition
of LDOS wiggles from several nearby impurities; it is necessary (and not too
difficult) for the analysis method to separate these contributions.}
\SAVE{In real materials, one can have 
extended scatterers or chains of point impurities and these will have their own signatures 
in an echo analysis. Analysing these will be a numerically-harder task and this will
form part of future work.}
\SAVE{Echoes rely on existence of well-defined Landau/BCS quasiparticles and thus mismatch of
this kind of analysis with experiment may point to features which are outside 
a quasiparticle description and need different physics to be described. }

\SAVE{
Given a material, one can imagine - as a starting
point - using FT-STS to figure out the dispersion of quasiparticles and then using 
echo analyses to investigate the disorder properties of 
the material. Further corroboration with other experimental methods would solidify
the understanding gained through STM.}

\SAVE{As a final point, we return to the ``inverse'' question
posed in the abstract : given an experimental data set,
how much can we infer about the microscopic Hamiltonian (and
how do we perform the inversion)?
Eliashberg theory \cite{Eliashberg, Scal_Parks} with the
MacMillan-Rowell inversion \cite{Rowell_Parks} was a prime example of this philosophy, whereby
tunneling data from metal-insulator-superconductor junctions yielded the electron-phonon 
coupling and the phonon density of states.
The vast amount of spatially dependent data that STM can capture motivates 
inversion to produce a spatially-dependent Hamiltonian representing the
particular sample measured. \OMIT{Analogous to MacMillan-Rowell Inversion,}
FT-STS and Echo analyses together provide an avenue to do just that for cuprates.
}

\acknowledgments
We thank S. C. Davis, A. V. Balatsky, and P. J. Hirschfeld for discussions.
S.P. thanks Stefan Baur for helping with Mathematica.
This work was supported by NSF Grant No. DMR-0552461.


\begin{thebibliography}{99}




\bibitem{Eigler} 
M. F. Crommie,  C. P. Lutz, and D. M. Eigler, Science 262, 218 (1993);
M. F. Crommie, C. P. Lutz, and D. M. Eigler, Nature (London) 363, 524 (1993).


%
%

\bibitem{Yazdani} A. Yazdani  {\it et al.}, Phys. Rev. Lett. 83, 176 (1999).

\bibitem{Hudson} E. W. Hudson {\it et al.},  Science 285, 88 (1999).

\bibitem{Pan} S. H. Pan, {\it et al}, Nature 413, 282 (2001).

\bibitem{Cren} T. Cren {\it et al}
Europhys. Lett. 54, 84 (2001).

\bibitem{Howald} 
C. Howald, P. Fournier, and A. Kapitulnik,  Phys. Rev. B 64, 100504 (2001).

\bibitem{Lang} 
K. M. Lang {\it et al.}, Nature 415, 412. 

\bibitem{Hoffman} 
J. E. Hoffman, {\it et al.}, Science 297, 1148
(2002).

\bibitem{Renner} 
O. Fischer {\it et al},
Rev. Mod. Phys. 79, 353 (2007).


%
%




\bibitem{Sprunger-Petersen} 
P.~T. Sprunger {\it et al.}, Science 275, 1764 (1997);
L. Petersen {\it et al.}, Phys. Rev. B 57, R6858 (1998).

\bibitem{wenderoth}
A. Weismann {\it et al},
Science 323, 1190 (2009).
These authors pinpoint scatterers, but using
spatial rather than energy oscillations.


%
%

\bibitem{WangLee-Ting} 
Q.-H. Wang and D.-H. Lee, Phys. Rev. B 67, 020511 (2003);
D. Zhang. and C. S. Ting, Phys. Rev. B 67, 020511 (2003)

\bibitem{McElroy-Hanaguri} K. McElroy {\it et al}, Nature 422, 592 (2003);
T. Hanaguri {\it et al.}, Nature Phys. 3, 865 (2007).

\bibitem{gutzwiller}
This is analogous to the oscillations in Gutzwiller's trace formula
in terms of the (classical) return times of a wavepacket.  See
M. C. Gutzwiller, {\it Chaos in Classical and Quantum Mechanics}
(Springer, New York, 1990).

\bibitem{McElroy-chem}
K. McElroy, {\it et al}
Science 309, 1048 (2005).


%
%
\bibitem{recursion} 
V. Heine, in: {\it Solid State Physics} vol. 35, eds. H. Ehrenreich, F.
Seitz and D. Turnbull (Academic Press, New York, 1980);
R. Haydock, in ibid.

\bibitem{Litak} 
G. Litak, P. Miller, B.L. Gyorffy, Physica C 251 (1995) 263-273.


\bibitem{FN-contour-sing}
{If the contour included a point with $v_g(\kkE(s))=0$, that 
point might also make a singular contribution.}

\bibitem{asymptotic}
N. Bleistein, and R. Handelsman, {\it Asymptotic Expansions of
Integrals} (Dover, New York, 1975).
%
\bibitem{11}
The (1,1) direction was chosen to bring out the difference in the two quasiparticle velocities most clearly.
Along other directions, there are two velocities
but they are very similar in magnitude \SAVE{(because the bananas are highly elongated/anisotropic
and hence the velocities are derived from near the tips in the two quadrants)} and, hence,
it is difficult to distinguish the two different wiggles. For this reason, possibly in
a real experiment, one might have to look for echoes in certain directions.


%

\bibitem{Balatsky_review} 
A. V. Balatsky et al., Rev. Mod. Phys. 78, 373 (2006).

\bibitem{Nunner} 
T. S. Nunner et al., Phys. Rev. B 73, 104511 (2006).


\bibitem{graphene-STM}
G. Li, A. Luican, and E. Y. Andrei,
Phys. Rev. Lett. 102, 176804 (2009).


\bibitem{Kohsaka} 
Y. Kohsaka et al., Nature 454, 1072-1078  (2008)

\bibitem{pointimp}
In this paper, we illustrated echolocation only for the case of {\it isolated}
point-like impurities. Real STM data will typically contain a superposition
of LDOS wiggles from several nearby impurities; it is necessary (and not too
difficult) for the analysis method to separate these contributions.


\end{thebibliography}
\end{document}